\documentclass[twocolumn]{autart}    

\usepackage{amssymb,amsmath,amsfonts}
\usepackage{algorithm,algorithmic}
\usepackage{cite}
\usepackage{float}
\usepackage{epsfig}
\usepackage{graphicx}
\usepackage{mathtools}

\usepackage{graphics}
\usepackage{multicol}
\usepackage{lipsum}
\usepackage{subfigure}
\usepackage{url}
\usepackage{verbatim}
\usepackage{xspace}
\usepackage[usenames,dvipsnames]{color}
\usepackage{setspace}
\usepackage{dblfloatfix}

\floatstyle{ruled}
\newfloat{model}{H}{mod}
\floatname{model}{\footnotesize Model}
\newfloat{notatio}{H}{not}
\floatname{notatio}{\footnotesize Notation}

\newenvironment{varalgorithm}[1]
  {\algorithm\renewcommand{\thealgorithm}{#1}}
  {\endalgorithm}

\newenvironment{list4}{
	\begin{list}{$\bullet$}{%
			\setlength{\itemsep}{0.05cm}
			\setlength{\labelsep}{0.2cm}
			\setlength{\labelwidth}{0.3cm}
			\setlength{\parsep}{0in} 
			\setlength{\parskip}{0in}
			\setlength{\topsep}{0in} 
			\setlength{\partopsep}{0in}
			\setlength{\leftmargin}{0.16in}}}
	{\end{list}}

\usepackage{url,changebar,bm,xspace,dsfont}
\let\mathbb=\mathds 

\newtheorem{theorem}{Theorem}

\newtheorem{example}{\bfseries Example}
\newtheorem{remark}{Remark}

\usepackage{graphics} 
\usepackage{epsfig} 
\usepackage{algorithm,algorithmic}

\usepackage{amsmath,amsfonts,amssymb,amscd}
\usepackage{capt-of}

\usepackage{color}
\usepackage{cite}
\usepackage{verbatim}

\newtheorem{lemma}{\bfseries Lemma}

\begin{document}

\begin{frontmatter}

\title{Finite Time Exact Quantized Average Consensus \\ with Limited Resources and Transmission Stopping for Energy-Aware Networks\thanksref{footnoteinfo}} 
 
\thanks[footnoteinfo]{The results of this paper were not presented at any conference.
Corresponding author Apostolos~I.~Rikos.}

\author[First]{Apostolos~I.~Rikos}\ead{rikos@kth.se}, 
\author[Second]{Christoforos~N.~Hadjicostis}\ead{chadjic@ucy.ac.cy}, 
\author[First]{Karl~H.~Johansson}\ead{kallej@kth.se} 

\address[First]{Division of Decision and Control Systems, KTH Royal Institute of Technology, \\ and also affiliated with Digital Futures, SE-100 44 Stockholm, Sweden}  
\address[Second]{Department of Electrical and Computer Engineering, University of Cyprus, Nicosia, Cyprus}             


\begin{abstract}                          
Composed of spatially distributed sensors and actuators that communicate through wireless networks, networked control systems are emerging as a fundamental infrastructure technology in $5$G and IoT technologies, including diverse applications, such as autonomous vehicles, UAVs, and various sensing devices.  
In order to increase flexibility and reduce deployment and maintenance costs, many such applications consider battery-powered or energy-harvesting networks, which bring additional limitations on the energy consumption of the wireless network. 
Specifically, the operation of battery-powered or energy-harvesting wireless communication networks needs to guarantee (i) efficient communication between nodes and (ii) preservation of available energy. 
Motivated by these novel requirements, in this paper, 
we present and analyze a novel distributed average consensus algorithm, which (i) operates exclusively on quantized values (in order to guarantee efficient communication and data storage), and (ii) relies on event-driven updates (in order to reduce energy consumption, communication bandwidth, network congestion, and/or processor usage). 
We characterize the properties of the proposed algorithm and show that its execution, on any time-invariant and strongly connected digraph, will allow all nodes to reach, in finite time, a common consensus value that is equal to the exact average (represented as the ratio of two quantized values). 
Furthermore, we show that our algorithm allows each node to cease transmissions once the exact average of the initial quantized values has been reached (in order to preserve its battery energy). 
Then, we present upper bounds on (i) the number of transmissions and computations each node has to perform during the execution of the algorithm, and (ii) the memory and energy requirements of each node in order for the algorithm to be executed. 
Finally, we provide examples that demonstrate the operation, performance, and potential advantages of our proposed algorithm. 
\end{abstract}


\end{frontmatter}

%
%
%
%
\section{INTRODUCTION}\label{intro}

In recent years there have been tremendous advances in the area of wireless networking, sensing, computing, and control. 
These advances are revolutionizing the role and importance of wireless control networks in various areas, such as Cyber-Physical Systems \cite{2012:Sztipanovits} and Internet of Things (IoT) applications \cite{2016:Bello_Zeadally}. 
Wireless control networks consist of various sensor nodes which sample and transmit data over a wireless channel to controllers, which are in charge of deciding the necessary actions based on the received data. 
Wireless networks play an important role in the rapid expansion of the areas of embedded computing, advanced control, cloud computing, and emerging applications in autonomous vehicles \cite{2015:Alam_Besselink_Martensson_Johansson, 2016:Demir_Ergen}, coordination of UAVs \cite{2016:Nowzari_Pappas}, and sensor networks \cite{2013:Aziz_Ivanovich}. 
A recent survey which analyzes the importance of wireless networks in emerging control systems can be found in \cite{2018:Park_Ergen_Fischione_Lu_Johansson}.

By their nature, wireless networks do not have a fixed infrastructure and do not use centralized methods for organization. 
This flexibility enables their use even when a fixed infrastructure is unavailable and makes them attractive for numerous applications (ranging from military, civil, industrial or environmental monitoring in hostile environments). 
The absence of cables for data communication motivates the removal of the power supply from the nodes in order to achieve even more flexibility. 
Therefore, nodes need to rely on (i) battery storage, and/or (ii) energy harvesting techniques for their operation.
Prolonging the lifetime of nodes and enhancing network flexibility through efficient battery management and/or energy harvesting techniques has received a lot of attention in recent years in the area of networked control systems \cite{2011:Park_Fischione_Bonivento_Johansson_Vincentelli, 2010:Ploennigs_Vasyutynskyy_Kabitzsch, 2019:Knorn_Dey_Ahlen_Quevedo, 2020:Ma_Lan_Hassan_Hu_Das, 2013:Aziz_Ivanovich}.

Battery-driven operation is of particular importance in cases where wireless networks are deployed in remote and inaccessible places (with no fixed infrastructure). 
However, modern wireless networks may consist of thousands of nodes and replacing/recharging batteries may be a costly, time-consuming or even infeasible task. 
Therefore, researchers have proposed several battery-driven energy conservation strategies to ensure energy efficient network operation. 
In \cite{2010:Quevedo_Ahlen_Ostergaard} the authors present a power and coding control algorithm for state estimation with wireless sensors. 
An online scheduling scheme was developed in \cite{2014:Han_Cheng_Chen_Shi} under communication energy constraints for remote state estimation. 
In \cite{2014:Gatsis_Ribeiro_Pappas} the authors study the control scheme of a linear plant when state information is being transmitted from a sensor to the controller over a wireless fading channel. 
Additionally, the graph routing problem optimized for maximizing network lifetime is analyzed in \cite{2016:Wu_Gunatilaka_Saifullah_Sha_Tiwari_Lu_Chen}.
Finally, self-triggered control determines its next execution time according to the triggering rule and previously received data \cite{2009:Wang_Lemmon} allowing for better allocation of network resources. 
Note, however, that battery-driven operation consists of limited energy supply constraints.

Energy constraints are widely regarded as a fundamental limitation of wireless devices. 
For this reason, many researchers aimed to develop alternative energy provision mechanisms by utilizing ambient energy. 
A sensor network can have near perpetual operation, by utilizing energy harvesting strategies in order to harvest ambient energy \cite{2020:Ma_Lan_Hassan_Hu_Das}. 
However, limitations on energy harvesting opportunities have led researchers to also work on the direction of efficient energy management in wireless sensor networks with energy harvesting capabilities. 
Specifically, power control policies for maximizing throughput or minimizing mean delay or transmission completion time were presented in \cite{2010:Sharma_Mukherji_Joseph_Gupta, 2012:Tutuncuoglu_Yener}, whereas power control algorithms for maximizing the mutual information of a wireless link were discussed in \cite{2012:Ho_Zhang}. 
Jointly controlling data queue and battery buffer in order to maximize the long-term average sensing rate was presented in \cite{2012:Mao_Koksal_Shroff}, while in \cite{2013:Nayyar_Basar_Teneketzis_Veeravalli} a sensor with energy harvesting capabilities that sends its measurements toward a remote estimator is considered. 
The design of optimal sensor transmission power control schemes was presented in \cite{2017:Li_Zhang_Quevedo_Lau_Dey_Shi}, and in \cite{2017:Du_Yang_Shen_Kwak} the distortion minimization problem by optimizing sleep-wake scheduling and transmit power was considered. 
In \cite{2019:Knorn_Dey_Ahlen_Quevedo}, the authors presented optimal transmission scheduling schemes in the presence of energy-leaking sensor batteries. 
The authors of \cite{2019:Bhat_Motani_Lim} improved throughput performance by considering circuit power and non-ideal batteries.
In \cite{2020:Vasconcelos_Gagrani_Nayyar_Mitra} an energy-harvesting scheduler is considered which makes independent observations and in \cite{2020:Nobar_Mansourkiaie_Ahmed} an optimization framework to minimize the summation of packet dropping probability is presented.
In \cite{2021:Seifullaev_Knorn_Ahlen} the authors construct and analyze a simplified sensor battery model that is used to predict sensor battery dynamics. 
However, in practice, the amount of harvested energy may be random and sometimes insufficient. 
Thus, energy efficient algorithms for wireless sensor networks still require further development.

In today's emerging technologies, the limited battery or harvested energy associated with wireless networks is a major bottleneck. 
This bottleneck may trammel the two main advantages of battery powered or energy harvesting wireless networks, which are (i) autonomy of the operating nodes, and (ii) network flexibility. 
Efficient usage of the limited energy resources is critically important to support these advantages and prolong the lifetime of the nodes. 
For this reason, there is an increasing demand for large scale network coordination algorithms which operate over battery powered networks or networks with energy harvesting capabilities, and whose efficient operation (i) reduces energy consumption, and (ii) achieves a lifetime of several years using nodes that carry merely hundreds of joules of stored/harvested energy.

In this paper, we focus on distributed control and coordination over wireless networks nodes that are battery powered or utilize energy harvesting techniques. 
We focus on the average consensus problem in which a group of nodes reaches agreement to a common value that is equal to the average of the initial states of the nodes \cite{2018:BOOK_Hadj, 2013:Themis_Hadj_Johansson}. 
Calculating the average of their initial states allows nodes to coordinate their actions via a common decision and is useful in many applications such as load balancing, voting schemes, quantized privacy protocols (can be used as the basis for various encoding schemes), and distributed optimization. 
However, in practical scenarios there exist constraints on the bandwidth of communication links and the capacity of physical memories. 
This means that network links can only allow messages of limited length (i.e., \textit{quantized}) to be stored and transmitted between nodes \cite{2016:Chamie_Basar, 2011:Cai_Ishii, 2020:Rikos_Quant_Cons, 2021:Rikos_Hadj_Splitting}. 
Furthermore, the desire to achieve more efficient usage of network resources has increased the interest in event-triggered algorithms for distributed coordination and, more generally, distributed control \cite{2013:Dimarogonas_Johansson, 2016:nowzari_cortes, 2018:Manitara_Hadj}. 
In most existing applications of wireless networks which consider a limited source of energy, their operation is not designed to guarantee efficient communication and energy preservation. 
Specifically, (i) nodes may operate in a way that is not ``event-based'' (i.e., nodes transmit their state at each iteration, which leads to increased energy consumption), or (ii) nodes do not have a distributed strategy to determine whether convergence has been achieved (thus they continue transmitting even after convergence has been achieved, which leads to increased energy and bandwidth consumption), or (iii) node operation considers the transmission of real-valued states (thus not guaranteeing efficient communication). 
This paper aims to fill this gap by proposing a distributed coordination algorithm that fulfills all above requirements. 
Specifically, we focus on three main strategies for slowing down the depletion of energy in wireless networks: (i) efficient communication, (ii) event-triggered operation, and (iii) transmission stopping. 
To the authors' knowledge, only \cite{2021:Rikos_TaskScheduling} presents an algorithm which allows nodes in a data center to coordinate and perform task allocation by exchanging quantized messages and eventually stopping their operation according to a distributed mechanism. 
However, the distributed stopping mechanism in \cite{2021:Rikos_TaskScheduling} requires knowledge of the digraph diameter which is a global parameter. 
Thus, the design of distributed coordination algorithms which (i) operate in an event-based fashion, (ii) consider efficient (quantized) communication, (iii) converge to the \textit{exact} quantized average of the initial states without any quantization error, and (iv) utilize a distributed stopping mechanism for ceasing transmissions without knowledge of global network parameters, is still an open question.

\subsection{Main Contributions}\label{Contributions}

In this paper, we present a novel distributed average consensus algorithm for battery powered or energy harvesting wireless networks, that combines the desirable features mentioned above. 
More specifically, average consensus is reached in finite time; processing, storing, and exchange of information between neighboring nodes is subject to uniform quantization; and the control actuation at each node is ``event-driven''. 
The main contributions of our paper are the following. 
\begin{itemize}
\item We present a novel distributed algorithm that is able to calculate the \textit{exact} average of the initial values in the form of a quantized fraction (i.e., as the ratio of two integer values) introducing no quantization error\footnote{Note that most algorithms in the available literature (see \cite{2007:Basar, 2009:Nedic, 2011:Cai_Ishii}) are able to converge to the ceiling or the floor of the initial average thus introducing a quantization error.}. 
\item We show that our algorithm converges to the desired result after a finite number of iterations, and we provide a polynomial upper bound on the number of time steps needed for convergence\footnote{The operation of our algorithm in this paper is analyzed over static directed graphs.
However, it can be extended also to dynamic networks which is a more suitable scenario for controlling UAV swarms and autonomous vehicles.}. 
\item We show that our algorithm utilizes its distributed stopping capability and transmissions are ceased for every node once the exact quantized average of the initial states is calculated. 
\item We calculate an upper bound on the number of transmissions and computations each node performs during the operation of our algorithm. 
\item We analyze the consumption of available resources by calculating an upper bound on the required memory and the required energy for each node during the operation of our algorithm. 
\item We demonstrate the operation of our algorithm via examples while we analyze its potential advantages and its transmission stopping capabilities. 
\end{itemize}
The operation of the proposed event-triggered algorithm essentially involves directed transmissions and broadcast transmissions from every node according to multiple event-triggered conditions. 
Specifically, every node broadcasts to every out-neighbor the value of its initial quantized state. 
This means that every node learns the maximum initial state in the network. 
Then, the nodes which have an initial state less than the maximum value transmit their quantized initial state directly to one neighboring node. 
The nodes that receive multiple directed messages from neighboring nodes sum the values, update their state and broadcast the updated state according to a set of event triggered conditions. 
The operation of the algorithm ensures that every initial quantized state is summed in a single node in the network. 
Then this node broadcasts its updated state (which is equal to the exact average of the initial quantized states) and every node in the network learns and updates its own state to be equal to the exact average. 
Once every node learns the state that is equal to the exact average convergence has been achieved and transmissions are ceased. 

Following \cite{2007:Basar, 2011:Cai_Ishii} we assume that the node's states are integer-valued (which comprises  a uniform class of quantization effects). 
Note that most work dealing with quantization has concentrated on the scenario where the nodes can store and process real-valued states but can transmit only quantized values through limited rate channels (see, \cite{2016:Chamie_Basar}). 
However, by contrast, our assumption is also suited to the case where the states are stored in digital memories of finite capacity (as in \cite{2009:Nedic, 2007:Basar, 2011:Cai_Ishii}), as long as the initial values are also quantized.

\subsection{Paper Organization}\label{Organization}

The remainder of this paper is organized as follows. 
In Section~\ref{preliminaries}, we introduce the notation used throughout the paper.
In Section~\ref{probForm} we formulate the finite transmission quantized average consensus problem. 
In Section~\ref{MaxAlgorithm}, we present a deterministic event-triggered distributed algorithm, which (i) allows the nodes to reach consensus to the \textit{exact} quantized average of the initial values after a finite number of steps, and (ii) allows them to cease transmissions once quantized average consensus is reached. 
Furthermore, we analyze the algorithm's operation via an example, and we calculate a worst case upper bound on the number of time steps required for convergence.  
In Section~\ref{bound_trans_comp}, we present a deterministic upper bound on the number of transmissions and computations each node performs during the operation of the algorithm. 
In Section~\ref{energy_constr}, we analyze the consumption of resources by calculating an upper bound on the required memory and the required energy of each node for the execution of the proposed algorithm. 
In Section~\ref{results}, we present simulation results and comparisons. 
We conclude in Section~\ref{future} with a brief summary and remarks about future work.

%
%
%
%
\section{NOTATION AND BACKGROUND}\label{preliminaries}

The sets of real, rational, integer and natural numbers are denoted by $ \mathbb{R}, \mathbb{Q}, \mathbb{Z}$ and $\mathbb{N}$, respectively. 
The symbol $\mathbb{Z}_+$ denotes the set of nonnegative integers.

\subsection{Graph-Theoretic Notions}

Consider a network of $n$ ($n \geq 2$) nodes communicating only with their immediate neighbors. 
The communication topology can be captured by a directed graph (digraph), called \textit{communication digraph}. 
A digraph is defined as $\mathcal{G}_d = (\mathcal{V}, \mathcal{E})$, where $\mathcal{V} =  \{v_1, v_2, \dots, v_n\}$ is the set of nodes and $\mathcal{E} \subseteq \mathcal{V} \times \mathcal{V} - \{ (v_j, v_j) \ | \ v_j \in \mathcal{V} \}$ is the set of edges (self-edges excluded). 
A directed edge from node $v_i$ to node $v_j$ is denoted by $m_{ji} \triangleq (v_j, v_i) \in \mathcal{E}$, and captures the fact that node $v_j$ can receive information from node $v_i$ (but not the other way around). 
We assume that the given digraph $\mathcal{G}_d = (\mathcal{V}, \mathcal{E})$ is \textit{strongly connected} (i.e., for each pair of nodes $v_j, v_i \in \mathcal{V}$, $v_j \neq v_i$, there exists a directed \textit{path}\footnote{A directed \textit{path} from $v_i$ to $v_j$ exists if we can find a sequence of vertices $v_i \equiv v_{l_0},v_{l_1}, \dots, v_{l_t} \equiv v_j$ such that $(v_{l_{\tau+1}},v_{l_{\tau}}) \in \mathcal{E}$ for $ \tau = 0, 1, \dots , t-1$.} from $v_i$ to $v_j$), which is the necessary (and sufficient) requirement for average consensus to be possible. 
The subset of nodes that can directly transmit information to node $v_j$ is called the set of in-neighbors of $v_j$ and is represented by $\mathcal{N}_j^- = \{ v_i \in \mathcal{V} \; | \; (v_j,v_i)\in \mathcal{E}\}$, while the subset of nodes that can directly receive information from node $v_j$ is called the set of out-neighbors of $v_j$ and is represented by $\mathcal{N}_j^+ = \{ v_l \in \mathcal{V} \; | \; (v_l,v_j)\in \mathcal{E}\}$. 
The cardinality of $\mathcal{N}_j^-$ is called the \textit{in-degree} of $v_j$ and is denoted by $\mathcal{D}_j^- = | \mathcal{N}_j^- |$, while the cardinality of $\mathcal{N}_j^+$ is called the \textit{out-degree} of $v_j$ and is denoted by $\mathcal{D}_j^+ = | \mathcal{N}_j^+ |$.

\subsection{Node Operation}

With respect to quantization of information flow, we have that at time step $k \in \mathbb{Z}_+$ (where $\mathbb{Z}_+$ is the set of nonnegative integers), each node $v_j \in \mathcal{V}$ maintains the transmission variables $S\_br_j \in \mathbb{N}$ and $M\_tr_j \in \mathbb{N}$, the state variables $y^s_j[k] \in \mathbb{Z}$, $z^s_j[k] \in \mathbb{Z}_+$ and $q_j^s[k] = \frac{y_j^s[k]}{z_j^s[k]}$, and the mass variables $y_j \in \mathbb{Z}$ and $z_j \in \mathbb{Z}_+$. 
Note here that for every node $v_j$ the transmission variables $S\_br_j$, $M\_tr_j$ are used to decide whether it will broadcast its state variables or transmit its mass variables, the state variables $y^s_j[k], z^s_j[k], q_j^s[k]$ are used to store the received messages and calculate the quantized average of the initial values, and the mass variables $y_j[k], z_j[k]$ are used to communicate with other nodes by either transmitting or receiving messages.

Furthermore, we assume that each node is aware of its out-neighbors and can directly transmit messages to each out-neighbor; however, it cannot necessarily receive messages (at least not directly) from them. 
In the proposed distributed protocol, each node $v_j$ assigns a \textit{unique order} in the set $\{0,1,..., \mathcal{D}_j^+ -1\}$ to each of its outgoing edges $m_{lj}$, where $v_l \in \mathcal{N}^+_j$. 
More specifically, the order of link $(v_l,v_j)$ for node $v_j$ is denoted by $P_{lj}$ (such that $\{P_{lj} \; | \; v_l \in \mathcal{N}^+_j\} = \{0,1,..., \mathcal{D}_j^+ -1\}$). 
This unique predetermined order is used during the execution of the proposed distributed algorithm as a way of allowing node $v_j$ to transmit messages to its out-neighbors in a \textit{round-robin}\footnote{When executing the proposed protocol, each node $v_j$ transmits to its out-neighbors, one at a time, by following a predetermined order. The next time it transmits to an out-neighbor, it continues from the outgoing edge it stopped the previous time and cycles through the edges in a round-robin fashion according to the predetermined ordering.} fashion.

\section{PROBLEM FORMULATION}\label{probForm}

Consider a strongly connected digraph $\mathcal{G}_d = (\mathcal{V}, \mathcal{E})$, where each node $v_j \in \mathcal{V}$ has an initial (i.e., for $k=0$) quantized value $y_j[0]$ (for simplicity, we take $y_j[0] \in \mathbb{Z}$). 
In this paper, we develop a distributed algorithm that allows nodes to address the problem presented below, while processing and transmitting \textit{quantized} information via available communication links. 

Each node $v_j$ obtains, after a finite number of steps, a fraction $q^s$ which is equal to the \textit{exact} average $q$ of the initial values of the nodes (i.e., there is no quantization error), where
\begin{equation}
q = \frac{\sum_{l=1}^{n}{y_l[0]}}{n} .
\end{equation}
Specifically, we argue that there exists $k_0$ so that for every $k \geq k_0$ we have 
\begin{equation}\label{alpha_z_y}
y^s_j[k] = \frac{\sum_{l=1}^{n}{y_l[0]}}{\alpha}  \ \ \text{and} \ \ z^s_j[k] = \frac{n}{\alpha} ,
\end{equation}
where $\alpha \in \mathbb{N}$. This means that 
\begin{equation}\label{alpha_q}
q^s_j[k] = \frac{(\sum_{l=1}^{n}{y_l[0]}) / \alpha}{n / \alpha} \coloneqq q ,
\end{equation}
for every $v_j \in \mathcal{V}$ (i.e., for $k \geq k_0$ every node $v_j$ has calculated $q$ as the ratio of two integer values). 
Furthermore, we have that every node $v_j$ stops performing transmissions towards its out-neighbors $v_l \in \mathcal{N}^+_j$ once its state variables $y^s_j$, $z^s_j$, $q^s_j$ fulfill \eqref{alpha_z_y} and \eqref{alpha_q}, respectively.


%
%
%
%
\section{EVENT-TRIGGERED QUANTIZED AVERAGE CONSENSUS ALGORITHM WITH FINITE TRANSMISSION CAPABILITIES}\label{MaxAlgorithm}

In this section we present a distributed algorithm which achieves \textit{exact} quantized average consensus in a finite number of time steps. 
Also, once average consensus is reached, all transmissions are ceased. 
The main idea is to maintain a separate mechanism for broadcasting the state variables and the mass variables of each node (as long as they satisfy certain event trigger conditions). 
This way, nodes learn the average but also have a way to decide when (or not) to transmit.

\subsection{Finite Transmission Event-Triggered Algorithm}

The details of the distributed algorithm with transmission stopping capabilities can be seen in Algorithm~\ref{algorithm_max}.

\noindent
\vspace{-0.5cm}    
\begin{varalgorithm}{1}
\caption{Finite Transmission Event-Triggered Quantized Average Consensus}
\textbf{Input:} A strongly connected digraph $\mathcal{G}_d = (\mathcal{V}, \mathcal{E})$ with $n=|\mathcal{V}|$ nodes and $m=|\mathcal{E}|$ edges. Each node $v_j\in \mathcal{V}$ has an initial state $y_j[0] \in \mathbb{Z}$. 
\\
\textbf{Initialization:} Each node $v_j \in \mathcal{V}$ does the following: 
\begin{list4}
\item[1)] Assigns to each outgoing edge $v_l \in \mathcal{N}^+_j$ a \textit{unique order} $P_{lj}$ in the set $\{0,1,..., \mathcal{D}_j^+ -1\}$.
\item[2)] Sets $z_j[0] = 1$, $z^s_j[0] = 1$, $y^s_j[0] = y_j[0]$, $q^s_j[0] = y^s_j[0] / z^s_j[0]$ and $S\_br_j = 0$, $M\_tr_j = 0$. 
\item[3)] Broadcasts $z^s_j[0]$, $y^s_j[0]$ to every $v_l \in \mathcal{N}_j^+$.
\end{list4}
\textbf{Iteration:} For $k=0,1,2,\dots$, each node $v_j \in \mathcal{V}$ does the following: 
\begin{list4}
\item[1)] Receives $y^s_i[k]$, $z^s_i[k]$ from every $v_i \in \mathcal{N}_j^-$ (if no message is received it sets $y^s_i[k] = 0$, $z^s_i[k] = 0$). 
\item[2)] Receives $y_i[k]$, $z_i[k]$ from each $v_i \in \mathcal{N}_j^-$ and sets 
$$
y_j[k+1] = y_j[k] + \sum_{v_i \in \mathcal{N}_j^-} w_{ji}[k]y_i[k] ,
$$  \vspace{-.3cm}
$$
z_j[k+1] = z_j[k] + \sum_{v_i \in \mathcal{N}_j^-} w_{ji}[k]z_i[k] ,
$$
where $w_{ji}[k]=1$ if a message with $y_i[k]$, $z_i[k]$ is received from in-neighbor $v_i$, otherwise $w_{ji}[k]=0$. 
\item[3)] \textbf{If} $w_{ji}[k] \neq 0$ or $z^s_i[k] \neq 0$ for some $v_i \in \mathcal{N}_j^-$ \textbf{then}
\begin{list4}
\item[3a)] Calls Algorithm~\ref{algorithm_max_1a}. 
\item[3b)] \textbf{If} $M\_tr_j = 1$ \textbf{then} chooses $v_l \in \mathcal{N}_j^+$ according to $P_{lj}$ (in a round-robin fashion) and transmits $y_j[k]$, $z_j[k]$. 
Then, sets $y_j[k] = 0$, $z_j[k] = 0$, $M\_tr_j = 0$. 
\item[3c)] \textbf{If} $S\_br_j = 1$ \textbf{then} broadcasts $z^s_j[k+1]$, $y^s_j[k+1]$ to every $v_l \in \mathcal{N}_j^+$. 
Then, sets $S\_br_j = 0$. 
\end{list4}
\item[4)] Repeats (increases $k$ to $k + 1$ and goes back to Step~$1$).
\end{list4}
\textbf{Output:} \eqref{alpha_q} holds for every $v_j \in \mathcal{V}$. 
\label{algorithm_max}
\end{varalgorithm}

\noindent
\vspace{-0.5cm}    
\begin{varalgorithm}{1.A}
\caption{Event-Triggered Conditions for Algorithm~\ref{algorithm_max} (for each node $v_j$)}
\textbf{Input} 
\\ $y^s_j[k]$, $z^s_j[k]$, $q^s_j[k]$, $y_j[k+1]$, $z_j[k+1]$, $S\_br_j$, $M\_tr_j$ and the received $y^s_i[k]$, $z^s_i[k]$ from every $v_i \in \mathcal{N}_j^-$.
\\
\textbf{Execution} 
\begin{list4}
\item[1)] \underline{Event Trigger Conditions~$1$:} \textbf{If} 
\\ Condition~$(i)$: $z^s_i[k] > z^s_j[k]$, or
\\ Condition~$(ii)$: $z^s_i[k] = z^s_j[k]$ and $y^s_i[k] > y^s_j[k]$, 
\\ \textbf{then} sets 
$$ 
z^s_j[k+1] = \max_{v_i \in \mathcal{N}_j^-} z^s_i[k] , \ \ \text{and}
$$ 
$$ 
y^s_j[k+1] = \max_{v_i \in \{v_{i'} \in \mathcal{N}_j^- | z^s_{i'}[k] = z^s_j[k+1]\}} y^s_i[k] ,
$$ 
and sets $q^s_j[k+1] = \frac{y^s_j[k+1]}{z^s_j[k+1]}$, and $S\_br_j = 1$. 
\item[2)] \underline{Event Trigger Conditions~$2$:} \textbf{If}
\\ Condition~$(i)$: $z_j[k+1] > z^s_j[k+1]$, or 
\\ Condition~$(ii)$: $z_j[k+1] = z^s_j[k+1]$ and $y_j[k+1] > y^s_j[k+1]$, 
\\ \textbf{then} sets $z^s_j[k+1] = z_j[k+1]$, $y^s_j[k+1] = y_j[k+1]$ and sets $q^s_j[k+1] = \frac{y^s_j[k+1]}{z^s_j[k+1]}$ and $S\_br_j = 1$.
\item[3)] \underline{Event Trigger Conditions~$3$:} \textbf{If}
\\ Condition~$(i)$: $0 < z_j[k+1] < z^s_j[k+1]$ or 
\\ Condition~$(ii)$: $z_j[k+1] = z^s_j[k+1]$ and $y_j[k+1] < y^s_j[k+1]$, 
\\ \textbf{then} sets $M\_tr_j = 1$.
\end{list4}
\textbf{Output} 
\\ $y^s_j[k]$, $z^s_j[k]$, $q^s_j[k]$, $S\_br_j$, $M\_tr_j$.
\label{algorithm_max_1a}
\end{varalgorithm}

\vspace{.1cm}

The intuition behind Algorithm~\ref{algorithm_max} is the following. 
Let us first consider the notion of ``leading mass''.
During time step $k$, the node that holds the pair of mass variables with the largest $z[k]$ value is referred to as the ``leading mass'' (pair). 
In case there are multiple nodes with pairs of mass variables that have the largest $z[k]$, then the ``leading mass'' is the pair (or pairs) of mass variables that has (or have) the largest $y[k]$ value among the pairs of mass variables with the largest $z[k]$. 
Note that a formal definition of the ``leading mass'' is presented in Subsection~\ref{sec:Conv_analysis}. 
During the Initialization process, each node $v_j$ considers its set of stored mass variables to be the ``leading mass''. 
For this reason, it sets its state variables to be equal to the stored mass variables, and then broadcasts the values of its state variables. 
During the Iteration process, each node $v_j$ (i) receives any (possibly) transmitted set of state variables from its in-neighbors and, (ii) receives and stores any (possibly) transmitted set of mass variables from its in-neighbors. 
If it received a set of state variables and/or a set of mass variables from its in-neighbors, then it executes Algorithm~\ref{algorithm_max_1a}. 
During the execution of Algorithm~\ref{algorithm_max_1a}, each node checks (i) Event Trigger Conditions~$1$, (ii) Event Trigger Conditions~$2$, and (iii) Event Trigger Conditions~$3$. 
In Event Trigger Conditions~$1$, it checks whether the received set of state variables is equal to the ``leading mass'' (in case it receives messages from multiple in-neighbors it checks which set of state variables is the ``leading mass''). 
If Event Trigger Conditions~$1$ hold, it sets its state variables to be equal to the received set of state variables which is the ``leading mass'' and decides to broadcast its updated state variables (i.e., sets its transmission variable $S\_br_j = 1$). 
In Event Trigger Conditions~$2$, it checks whether the set of mass variables it stored is the ``leading mass''. 
If Event Trigger Conditions~$2$ hold, it sets its state variables to be equal to the stored set of mass variables and decides to broadcast its updated state variables (i.e., sets its transmission variable $S\_br_j = 1$). 
In Event Trigger Conditions~$3$, it checks whether the set of mass variables it stored is not the ``leading mass''. 
Specifically, it checks whether its state variables are equal to the ``leading mass''.  
If Event Trigger Conditions~$3$ hold, this means that the mass variables of another node in the network is the ``leading mass'' (and the state variables of node $v_j$ became equal to the ``leading mass'' from Event Trigger Conditions~$1$). 
This means the stored mass variables is not the ``leading mass'' and thus $v_j$ decides to transmit its stored mass variables (i.e., sets its transmission variable $M\_tr_j = 1$). 
Once Algorithm~\ref{algorithm_max_1a} is executed, $v_j$ returns to the Iteration process of Algorithm~\ref{algorithm_max} and broadcasts its state variables and/or transmits its mass variables according to its transmission variables.  
Then, it repeats the procedure.

\begin{remark}
Notice here that each node $v_j$, during time step $k$, is able to perform two types of transmissions towards its out-neighbors $v_l \in \mathcal{N}_j^+$. 
It can broadcast (to all of its out-neighbors) its state variables $y^s_j[k]$ and $z^s_j[k]$ (if Event Trigger Conditions~$1$ and/or Event Trigger Conditions~$2$ hold) and it can transmit its mass variables $y_j[k]$ and $z_j[k]$ to a single out-neighbor, chosen according to the predetermined order $P_{lj}$ (if Event Trigger Conditions~$3$ hold). 
This may seem as a departure from the literature on average consensus which assumes only a broadcast primitive (see \cite{2010:christoforos, 2011:Christoforos-Themis, 2011:Franceschelli} and references therein) and the literature on quantized average consensus which assumes only a unicast primitive (i.e., directed transmissions) \cite{2007:Basar, 2011:Cai_Ishii, 2020:Rikos_Quant_Cons, 2018:RikosHadj_CDC}, or a broadcast primitive \cite{2016:Chamie_Basar}. 
However, with broadcast as sole primitive and without additional assumptions, achieving the exact average (i.e., avoiding an error introduced due to quantization) would be impossible (see \cite{2015:Hendrickx_Tsitsiklis}) while with unicast as a sole primitive exhibiting distributed stopping capabilities in order to terminate transmissions also appears difficult (e.g., see \cite{2020:Rikos_Quant_Cons} where the number of  transmissions at each time step monotonically decreases but it never becomes equal to zero). 
For this reason, each node $v_j$ is allowed to perform both types of transmissions (broadcast and unicast) and, as we will also see later, this achieves both exact average and transmission termination during the operation of Algorithm~\ref{algorithm_max}. 
This is a direct result of Event Trigger Conditions~$1$, $2$ and $3$ that characterize the operation of our algorithm and they effectively imply that no transmission is performed if no set of conditions holds when using Algorithm~\ref{algorithm_max_1a} to check them. 

It is also important to note that Algorithm~\ref{algorithm_max} can be applied to the standard average consensus problem, where the initial value of each node and the transmitted messages are real values. 
In this case, our proposed protocol allows deterministic convergence to the {\em exact} value after a finite number of time steps. 
This is an important aspect as most finite time algorithms are only able to calculate the average of the initial values within an error bound (e.g., see \cite{2016:Manitara_Hadj} and references therein) which is a direct consequence of their asymptotic convergence. $\hfill \blacksquare$
\end{remark}

During the operation of Algorithm~\ref{algorithm_max}, nodes are able to reach quantized average consensus after a finite number of steps. 
Depending on the graph structure and the initial mass variables of each node, we have the following two possible scenarios:
\begin{enumerate}
\item[A.] Full Mass Summation (i.e., there exists $k'_0 \in \mathbb{Z}_+$ where we have $y_j[k'_0] =\sum_{l=1}^{n}{y_l[0]}  \ \ \text{and} \ \ z_j[k'_0] = n$, for some node $v_j \in \mathcal{V}$, and $y_i[k'_0] = 0  \ \ \text{and} \ \ z_i[k'_0] = 0$, for each $ v_i \in \mathcal{V} - \{ v_j \}$). 
In this scenario (\ref{alpha_z_y}) and (\ref{alpha_q}) hold (eventually, for some $k_0 > k_0'$) for each node $v_j$ for the case where $\alpha = 1$. 
\item[B.] Partial Mass Summation (i.e., there exists $k'_0 \in \mathbb{Z}_+$ so that for every $k \geq k'_0$ there exists a set $\mathcal{V}^p[k] \subseteq \mathcal{V}$ in which we have $y_j[k] = y_i[k]$ and $z_j[k] = z_i[k]$, $\forall v_j, v_i \in \mathcal{V}^p[k]$ and $y_l[k] = 0  \ \ \text{and} \ \ z_l[k] = 0$, for each $ v_l \in \mathcal{V} - \mathcal{V}^p[k]$). 
In this scenario, (\ref{alpha_z_y}) and (\ref{alpha_q}) hold (eventually, for some $k_0 > k_0'$) for each node $v_j$ for the case where $\alpha = | \mathcal{V}^p[k] |$.
\end{enumerate}

We now illustrate the event-triggered behavior of the proposed distributed algorithm via an example where we have ``Partial Mass Summation''.

\begin{example}
Consider a strongly connected digraph $\mathcal{G}_d = (\mathcal{V}, \mathcal{E})$, shown in Fig.~\ref{max_example}, with $\mathcal{V} = \{ v_1, v_2, v_3, v_4 \}$ and $\mathcal{E} = \{ m_{31}, m_{41}, m_{12}, m_{13}, m_{43}, m_{24} \}$ where each node has an initial quantized value $y_1[0] = 2$, $y_2[0] = 4$, $y_3[0] = 7$ and $y_4[0] = 9$ respectively. 
The average of the initial values, is equal to $q = \frac{22}{4}$. 

\begin{figure}[h]
\begin{center}
\includegraphics[width=0.3\columnwidth]{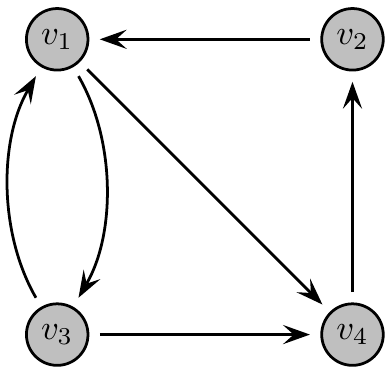}
\caption{Example of digraph for partial mass summation when using Algorithm~\ref{algorithm_max}.}
\label{max_example}
\end{center}
\end{figure}

Each node $v_j \in \mathcal{V}$ follows the Initialization steps in Algorithm~\ref{algorithm_max}. 
This means that it assigns to each of its outgoing edges $v_l \in \mathcal{N}^+_j$ a unique order $P_{lj}$ in the set $\{0,1,..., \mathcal{D}_j^+ -1\}$. 
Assume that the unique orders assigned by each node are the following:
$
v_1: P_{41} = 0, P_{31} = 1; \ v_2: P_{12} = 0; \ v_3: P_{13} = 0, P_{43} = 1, \ v_4: P_{24} = 0.  
$
For example, node $v_1$ will first transmit towards node $v_4$ and then transmit towards node $v_3$. 
Furthermore, according to the Initialization steps, each node $v_j$ sets its transmission variables $S\_br_j = 0$, $M\_tr_j = 0$ and then it broadcasts its state variables $z^s_j[0]$ and $y^s_j[0]$ to every out-neighbor $v_l \in \mathcal{N}_j^+$.
The initial mass and state variables, at time step $k=0$, for nodes $v_1, v_2, v_3, v_4$ are shown in Table~\ref{table_max1}. 

During the operation of Algorithm~\ref{algorithm_max}, at time step $k=0$ each node $v_j$ will receive the state variables $z^s_i[0]$ and $y^s_i[0]$ from every in-neighbor $v_i \in \mathcal{N}_j^-$ (however, it will not receive any set of mass variables $z_i[0]$ and $y_i[0]$ from any in-neighbor $v_i \in \mathcal{N}_j^-$). 
According to the Event Trigger Conditions~$1$, nodes $v_1$ and $v_2$ will update their state variables and then set $S\_br_1 = 1$, $S\_br_2 = 1$ (nodes $v_3$ and $v_4$ will maintain $S\_br_3 = 0$ and $S\_br_4 = 0$ since Event Trigger Conditions~$1$ do not hold for them). 
Furthermore, Event Trigger Conditions~$2$ do not hold for any node but Event Trigger Conditions~$3$ hold for nodes $v_1$ and $v_2$ who set $M\_tr_1 = 1$ and $M\_tr_2 = 1$ (nodes $v_3$ and $v_4$ will maintain $M\_tr_3 = 0$ and $M\_tr_4 = 0$ since Event Trigger Conditions~$3$ do not hold for them). 
Then nodes $v_1$ and $v_2$ will broadcast their state variables to every out-neighbor (since $S\_br_1 = 1$, $S\_br_2 = 1$) and then they will transmit their mass variables according to their unique predetermined order (since $M\_tr_1 = 1$ and $M\_tr_2 = 1$). 
Furthermore, they will set $S\_br_1 = 0$, $M\_tr_1 = 0$ and $S\_br_2 = 0$, $M\_tr_2 = 0$ since they transmitted their state and mass variables. 
The mass and state variables, at time step $k=1$, for nodes $v_1, v_2, v_3, v_4$ are shown in Table~\ref{table_max1}.

During time step $k=1$, each node $v_j$ will receive the state variables and the mass variables from every in-neighbor. 
Specifically, node $v_1$ will receive the mass variables of node $v_2$ and node $v_4$ will receive the mass variables of node $v_1$. 
Following Event Trigger Conditions~$1$, $v_1$ will update its state variables and set $S\_br_1 = 1$. 
Then, following Event Trigger Conditions~$2$, node $v_4$ will update its mass variables and set $S\_br_4 = 1$. 
Following Event Trigger Conditions~$3$, node $v_1$ will set $M\_tr_1 = 1$. 
Nodes $v_1$ and $v_4$ will broadcast their state variables to every out-neighbor (since $S\_br_1 = 1$, $S\_br_4 = 1$) and node $v_1$ will transmit its mass variables according to its unique predetermined order towards node $v_3$ (since $M\_tr_1 = 1$). 
Then, nodes $v_1$, $v_4$ will set $S\_br_1 = 0$, $S\_br_4 = 0$ and node $v_1$ will set $M\_tr_1 = 0$. 
The mass and state variables, at time step $k=2$, for nodes $v_1, v_2, v_3, v_4$ are shown in Table~\ref{table_max1}.

Repeating the steps above, in Table~\ref{table_max1} we can see the mass and state variables, at time step $k=3$. 
In Table~\ref{table_max1} we can see that for the set $\mathcal{V}^p[3] = \{v_3, v_4\}$ we have $y_3[3] = y_4[3]$ and $z_3[3] = z_4[3]$ while, for the set $ \mathcal{V} - \mathcal{V}^p[3] = \{v_1, v_2\}$ we have $y_1[3] = y_2[3] = 0$ and $z_1[3] = z_2[3] = 0$. 
This means that we have a ``Partial Mass Summation'' scenario. 
In this case, we will see that Event Trigger Conditions~$2$ will not hold again for any node for time steps $k > 3$ (i.e., no node will transmit again its mass variables).

\begin{table}[t]
\begin{center}
\captionof{table}{Mass and State Variables for Fig.~\ref{max_example}.}
\label{table_max1}
{\small 
\begin{tabular}{|c||c|c|c|c|c|}
\hline
Node &\multicolumn{5}{c|}{Mass and State Variables for $k=0$}\\
 &$y_j[0]$&$z_j[0]$&$y^s_j[0]$&$z^s_j[0]$&$q^s_j[0]$\\
\cline{1-6}
 &  &  &  &  & \\
$v_1$ & 2 & 1 & 2 & 1 & 2 / 1\\
$v_2$ & 4 & 1 & 4 & 1 & 4 / 1\\
$v_3$ & 7 & 1 & 7 & 1 & 7 / 1\\
$v_4$ & 9 & 1 & 9 & 1 & 9 / 1\\
\hline
Node &\multicolumn{5}{c|}{Mass and State Variables for $k=1$}\\
 &$y_j[1]$&$z_j[1]$&$y^s_j[1]$&$z^s_j[1]$&$q^s_j[1]$\\
\cline{1-6}
 &  &  &  &  & \\
$v_1$ & 2 & 1 & 7 & 1 & 7 / 1\\
$v_2$ & 4 & 1 & 9 & 1 & 9 / 1\\
$v_3$ & 7 & 1 & 7 & 1 & 7 / 1\\
$v_4$ & 9 & 1 & 9 & 1 & 9 / 1\\
\hline
Node &\multicolumn{5}{c|}{Mass and State Variables for $k=2$}\\
 &$y_j[2]$&$z_j[2]$&$y^s_j[2]$&$z^s_j[2]$&$q^s_j[2]$\\
\cline{1-6}
 &  &  &  &  & \\
$v_1$ & 4 & 1 & 9 & 1 & 9 / 1\\
$v_2$ & 0 & 0 & 9 & 1 & 9 / 1\\
$v_3$ & 7 & 1 & 7 & 1 & 7 / 1\\
$v_4$ & 11 & 2 & 11 & 2 & 11 / 2\\
\hline
Node &\multicolumn{5}{c|}{Mass and State Variables for $k=3$}\\
 &$y_j[3]$&$z_j[3]$&$y^s_j[3]$&$z^s_j[3]$&$q^s_j[3]$\\
\cline{1-6}
 &  &  &  &  & \\
$v_1$ & 0 & 0 & 9 & 1 & 9 / 1\\
$v_2$ & 0 & 0 & 11 & 2 & 11 / 2\\
$v_3$ & 11 & 2 & 11 & 2 & 11 / 2\\
$v_4$ & 11 & 2 & 11 & 2 & 11 / 2\\
\hline
Node &\multicolumn{5}{c|}{Mass and State Variables for $k=4$}\\
 &$y_j[4]$&$z_j[4]$&$y^s_j[4]$&$z^s_j[4]$&$q^s_j[4]$\\
\cline{1-6}
 &  &  &  &  & \\
$v_1$ & 0 & 0 & 11 & 2 & 11 / 2\\
$v_2$ & 0 & 0 & 11 & 2 & 11 / 2\\
$v_3$ & 11 & 2 & 11 & 2 & 11 / 2\\
$v_4$ & 11 & 2 & 11 & 2 & 11 / 2\\
\hline
\end{tabular}
}
\end{center}
\end{table}
\vspace{0.2cm}

During time step $k=4$, each node will receive the state variables and the mass variables from every in-neighbor. 
Node $v_1$ will update its state variables and set $S\_br_1 = 1$ (according to Event Trigger Conditions~$1$). 
Then, no transmissions of mass variables will be performed since Event Trigger Conditions~$3$ do not hold for any node; thus, for every node $v_j$ we have $M\_tr_j = 0$.
Node $v_1$ will broadcast its state variables to every out-neighbor (since $S\_br_1 = 1$) and every node $v_j$ will set $S\_br_j = 0$, $M\_tr_j = 0$. 
However, since the state variables of nodes $v_3$, $v_4$ are the same as the updated state variables of node $v_1$, we have the mass and state variables of every node at time step $k=5$ remain the same as time step $k=4$ (shown in Table~\ref{table_max1}).

In Table~\ref{table_max1}, we can see that (\ref{alpha_z_y}) and (\ref{alpha_q}) hold for every node for $\alpha = 2$ (i.e., every node has reached quantized average consensus). 
Notice that no set of event trigger conditions holds for any node during time step $k = 5$. 
This means that no node will perform any transmissions of its state or mass variables for time steps $k \geq 5$. \hspace*{\fill} $\square$
\end{example}

\begin{remark}
It is interesting to consider here the case where node $v_1$ sets its priorities as $P_{31} = 0$ and $P_{41} = 1$, during the Initialization of Algorithm~\ref{algorithm_max}. 
In this case, we will notice the scenario of ``Full Mass Summation'' for node $v_4$ instead of ``Partial Mass Summation'' (i.e., \eqref{alpha_z_y} and \eqref{alpha_q} hold for every node for $\alpha = 1$). $\hfill \blacksquare$
\end{remark}

\subsection{Deterministic Convergence Analysis}\label{sec:Conv_analysis}

We now analyze the functionality of Algorithm~\ref{algorithm_max} and prove that it allows all nodes to reach quantized average consensus after a finite number of steps. 
Furthermore, we will also show that once quantized average consensus is reached, transmissions from each node cease. 
We first consider the following setup and then state Lemma~\ref{before_second_lemma} and Lemma~\ref{second_lemma} which are necessary for our subsequent development.

{\it Setup:} Consider a strongly connected digraph $\mathcal{G}_d = (\mathcal{V}, \mathcal{E})$ with $n=|\mathcal{V}|$ nodes and $m=|\mathcal{E}|$ edges. 
During the execution of Algorithm~\ref{algorithm_max}, at time step $k_0$, there is at least one node $v_{j'} \in \mathcal{V}$, for which 
\begin{equation}\label{great_z_prop1_det}
z_{j'}[k_0] \geq z_i[k_0], \ \forall v_i \in \mathcal{V}.
\end{equation}
Then, among the nodes $v_{j'}$ for which (\ref{great_z_prop1_det}) holds, there is at least one node $v_j$ for which 
\begin{equation}\label{great_z_prop2_det}
y_j[k_0] \geq y_{l}[k_0] , \ \text{where} \ \ v_j, v_{l} \in \{ v_{j'} \in \mathcal{V} \ | \ (\ref{great_z_prop1_det}) \ \text{holds} \}.
\end{equation}
For notational convenience we will call the pair of mass variables of node $v_j$ for which (\ref{great_z_prop1_det}) and (\ref{great_z_prop2_det}) hold as the ``leading mass'' (or ``leading masses'' if multiple nodes hold such a pair of values) and the pairs of mass variables of a node $v_l$ for which $z_l[k_0]>0$ but (\ref{great_z_prop1_det}) and (\ref{great_z_prop2_det}) do not hold as the ``follower mass'' (or ``follower masses'').
Furthermore, if two (or more) masses reach a node simultaneously then we say that they ``merge'', i.e., the receiving node ``merges'' the mass variables it receives by summing their numerators and their denominators (according to Step~$2$ of the Iteration of Algorithm~\ref{algorithm_max}). 
This way a set of mass variables with a greater denominator is created.

\begin{lemma}\label{before_second_lemma}
If, during time step $k_0$ of Algorithm~\ref{algorithm_max}, the mass variables of node $v_j$ fulfill (\ref{great_z_prop1_det}) and (\ref{great_z_prop2_det}), then the state variables of every node $v_i \in \mathcal{V}$ satisfy 
\begin{eqnarray}\label{first_z}
z_i^s[k_0] \leq z_j[k_0] ,
\end{eqnarray}
or 
\begin{eqnarray}\label{first_zy}
z_i^s[k_0] = z_j[k_0] \ \ \text{and} \ \ y_i^s[k_0] \leq y_j[k_0].
\end{eqnarray}
\end{lemma}

\begin{pf}
Let us consider the variable
$
z^{(m)}[k] = \max_{v_l \in \mathcal{V}} z_l[k] . 
$
From Iteration Step~$2$ of Algorithm~\ref{algorithm_max} we have that $z^{(m)}[k]$ is non-decreasing (i.e., $z^{(m)}[k+1] \geq z^{(m)}[k]$, for every $k$). 
Furthermore, since the mass variables of node $v_j$ fulfill (\ref{great_z_prop1_det}) and (\ref{great_z_prop2_det}), then, during time step $k$, it holds that  
$
z_j[k] = z^{(m)}[k]. 
$
In addition, for every $k$, during Execution Steps~$1$ and $2$ of Algorithm~\ref{algorithm_max_1a}, for every node $v_i \in \mathcal{V}$, we have that $z_i^s[k]$ is either less than $z^{(m)}[k]$ (i.e., $z_i^s[k] < z^{(m)}[k]$) or equal to $z^{(m)}[k]$ (i.e., $z_i^s[k] = z^{(m)}[k]$). 
This is a direct result of $z^{(m)}[k]$ being non-decreasing and Event Trigger Conditions~$1$ and $2$ where, at time step $k$, the state variables $z_i^s[k]$, $y_i^s[k]$ of a node $v_i$ are updated to be (i) either equal to $z_i[k]$, $y_i[k]$ if $z_i^s[k] > z_i[k]$ or $z_i^s[k] = z_i[k]$, $y_i^s[k] > y_i[k]$ or (ii) equal to $z_{i'}^s[k]$, $y_{i'}^s[k]$, $v_{i'} \in \mathcal{N}_i^-$ if $z_{i'}^s[k] > z_{i}^s[k]$ or $z_{i'}^s[k] = z_{i}^s[k]$, $y_{i'}^s[k] > y_{i}^s[k]$.
As a result, at time step $k$, the state variables of every node $v_i \in \mathcal{V}$ satisfy 
$
z_i^s[k] \leq z_j[k] .
$
Finally, from Execution Steps~$1$ and $2$ of Algorithm~\ref{algorithm_max_1a}, for every $k$, we have that if $z_i^s[k] = z_j[k]$ holds for some node $v_i$, then we have that either $y_i^s[k] < y_j[k]$ or $y_i^s[k] = y_j[k]$. 
[Note here that if $z_i^s[k] = z_j[k]$ and $y_i^s[k] > y_j[k]$, then the mass variables of $v_j$ do not fulfill (\ref{great_z_prop1_det}) and (\ref{great_z_prop2_det}) which is a contradiction.]
As a result we have that if the mass variables of node $v_j$ fulfill (\ref{great_z_prop1_det}) and (\ref{great_z_prop2_det}), then the state variables of every node $v_i \in \mathcal{V}$ satisfy (\ref{first_z}) or (\ref{first_zy}).  \hspace*{\fill} $\square$
\end{pf}

\begin{lemma}\label{second_lemma}
If, during time step $k_0$ of Algorithm~\ref{algorithm_max}, the mass variables of {\em each} node $v_j$ with nonzero mass variables fulfill (\ref{great_z_prop1_det}) and (\ref{great_z_prop2_det}), then we have only leading masses and no follower masses. 
This means that the Event Trigger Conditions~$3$ will never hold again for future time steps $k \geq k_0$. 
As a result, the transmissions that (may) take place will be only via broadcasting (from Event Trigger Conditions~$1$ and $2$) for at most $n-1$ time steps and then they will cease. 
\end{lemma}

\begin{pf}
Let us assume that during time step $k_0$ two (or more) mass variables merge at nodes $v_j$, $v_i$, so that these two nodes simultaneously become leading masses (more generally, we could have more than two leading masses) and all other nodes have zero mass variables. 
Since the mass variables of nodes $v_j$, $v_i$, during time step $k_0$, become leading masses then there exists a set $\mathcal{V}^p[k_0] \subseteq \mathcal{V}$ in which we have $y_j[k_0] = y_i[k_0]$ and $z_j[k_0] = z_i[k_0]$, $\forall v_j, v_i \in \mathcal{V}^p[k_0]$ and $y_l[k_0] = 0  \ \ \text{and} \ \ z_l[k_0] = 0$, for each $ v_l \in \mathcal{V} - \mathcal{V}^p[k_0]$.
Once this merge occurs then we have that, for both $v_j$ and $v_i$, the Event Trigger Conditions $1$ and the Event Trigger Conditions $3$ do not hold, but Event Trigger Conditions $2$ do hold. 
This means that $v_j$ and $v_i$ do not transmit their mass variables but rather they broadcast their new state variables to their out-neighbors. 
Then, their out-neighbors, $v_{l_j}$ and $v_{l_i}$ respectively, will update their state variables and broadcast their new state variables towards their out-neighbors. 
The updating and broadcasting of state variables will continue, until all nodes obtain state variables equal to $z^s_j[k_0]$ and $y^s_j[k_0]$. 
Note that during this update and broadcasting of state variables, no node transmits its mass variables. 
After at most $n-1$ steps, all nodes will be aware of the values $z^s_j[k_0]$ and $y^s_j[k_0]$, and at that point all transmissions will seize. \hspace*{\fill} $\square$
\end{pf}

\begin{theorem}
\label{PROP1_max}
The execution of Algorithm~\ref{algorithm_max} allows each node $v_j \in \mathcal{V}$ to reach quantized average consensus after a finite number of steps $k_0$ upper bounded by $n^2 + (n-1)m^2$, where $n$ is the number of nodes and $m$ is the number of edges in the network. 
Furthermore, each node stops transmitting towards its out-neighbors once quantized average consensus is reached. 
\end{theorem}

\begin{pf}
Before starting the analysis of Algorithm~\ref{algorithm_max}, it is important to note that the leading mass will not fulfill the Event Trigger Conditions~$3$ in Execution Step~$3$ of Algorithm~\ref{algorithm_max_1a}. 
This means that the corresponding node (say $v_j$) will not transmit its mass variables to its out-neighbors $v_l \in \mathcal{N}_j^+$ according to its predetermined priority. 
In this proof we will show that there exists $k_0 \in \mathbb{Z}_+$ for which the mass variables of {\em every} node $v_j$ (for which $z_j[k] > 0$) fulfill (\ref{great_z_prop1_det}) and (\ref{great_z_prop2_det}), for every $k \geq k_0$. 
This means that for $k \geq k_0$ we have only leading masses. 
Furthermore, from Lemma~\ref{second_lemma}, we have that there exists $k_1 > k_0$, where for every $k \geq k_1$ the state variables of every node $v_j \in \mathcal{V}$ fulfill (\ref{alpha_z_y}) and (\ref{alpha_q}) for $\alpha \in \mathbb{Z}_+$ (i.e., every node has reached quantized consensus) and thus transmissions cease.

During the Initialization steps of Algorithm~\ref{algorithm_max}, we have that each node will broadcast its state variables to every out-neighbor. 
Then, during Iteration Step~$1$, each node will receive and update its state variables. 
When checking Event Trigger Conditions~$1$ it will decide to broadcast towards its out-neighbors the updated values (of the state variables). 
This means that after $n$ iterations (assuming, that no other mass variables merged during $n$ time steps), the state variables of each node $v_i \in \mathcal{V}$ satisfy $z^s_i[n] = z_{j_1}[0]$ and $y^s_i[n] = y_{j_1}[0]$, 
where the mass variables of node $v_{j_1}$ are the leading mass. 
As a result, after $n$ iterations, we have that the Event Trigger Conditions~$3$ will hold for every node $v_i \in \mathcal{V} - \{ v_{j_1} \}$. 
Thus, every node (except node $v_{j_1}$ which is the leading mass) will transmit its mass variables toward its out-neighbors according to its unique priority. 
Note here that the number of iterations required for the follower mass to reach every node $v_i \in \mathcal{V}$ is bounded by $m^2$, where $m = | \mathcal{E} |$ is the number of edges of the given digraph $\mathcal{G}_d$ (in this case Proposition~$3$ in \cite{2014:RikosHadj} provides a bound for the follower mass to travel via each edge in the graph and thus necessarily also reach every other node).
Let us assume now that, after executing Algorithm~\ref{algorithm_max} for additional $m^2$ steps, we have that the mass variables $z_{i_1}[0]$, $y_{i_1}[0]$ and $z_{i_2}[0]$, $y_{i_2}[0]$ of nodes $v_{i_1}$ and $v_{i_2}$ respectively, meet (and merge) in node $v_{j_2}$, and after this merge they become the leading mass.  
This means that node $v_{j_2}$ will not transmit its mass variables during time step $n + m^2$ (because Event Trigger Conditions~$3$ do not hold) but it will broadcast its state variables to every out-neighbor (because Event Trigger Conditions~$2$ hold). 
Thus, after additional $n$ iterations, the state variables of each node $v_i \in \mathcal{V}$ satisfy $z^s_i[2n + m^2] = z_{j_2}[n + m^2]$ and $y^s_i[2n + m^2] = y_{j_2}[n + m^2]$, where the mass variables of node $v_{j_2}$ are now the leading mass. 
This means that Event Trigger Conditions~$3$ will hold for every node $v_i \in \mathcal{V} - \{ v_{j_2} \}$. 
Thus, every node (except node $v_{j_2}$ which is now the leading mass) will transmit its mass variables toward its out-neighbors according to its unique priority. 
Note here that also node $v_{j_1}$ will transmit its mass variables toward its out-neighbors (since the state variables of $v_{j_1}$ are equal to the mass variables of the leading mass $v_{j_2}$, this means that Event Trigger Conditions~$3$ will also hold for $v_{j_1}$). 
Let us assume now that, after executing Algorithm~\ref{algorithm_max} for additional $m^2$ steps, the mass variables $z_{i_3}[0]$, $y_{i_3}[0]$ and $z_{i_4}[0]$, $y_{i_4}[0]$ of nodes $v_{i_3}$ and $v_{i_4}$ respectively, meet (and merge) in node $v_{j_3}$.
After this merge they become the leading mass. 
Again, this means that during time step $2n + 2m^2$, node $v_{j_3}$ will not transmit its mass variables (because Event Trigger Conditions~$3$ do not hold) but it will broadcast its state variables to every out-neighbor (because Event Trigger Conditions~$2$ hold). 
After additional $n$ iterations, the state variables of each node $v_i \in \mathcal{V}$ satisfy $z^s_i[3n + 2m^2] = z_{j_3}[2n + 2m^2]$ and $y^s_i[3n + 2m^2] = y_{j_3}[2n + 2m^2]$, where the mass variables of node $v_{j_3}$ are now the new leading mass. 
By continuing this analysis, we can see that every $n + m^2$ time steps at least two follower masses merge and become the leading mass. 
Since, during the Initialization steps of Algorithm~\ref{algorithm_max}, we have $n$ initial mass variables this means that after $(n - 1)(n + m^2)$ time steps {\it all} initial mass variables will merge into one mass (obviously the mass variables in which every initial mass has merged is the leading mass). 
Thus, at time step $(n-1)n + (n-1)m^2$ we have only leading masses and no follower masses. 
From Lemma~\ref{second_lemma}, we have that after $n$ additional time steps every node will have state variables equal to the leading mass (i.e., $z^s_i[n^2 + (n-1)m^2] = n$ and $y^s_i[n^2 + (n-1)m^2] = \sum_{l=1}^{n}{y_l[0]}$, for every $v_i \in \mathcal{V}$).  
As a result, each node $v_j \in \mathcal{V}$ will reach quantized average consensus after a finite number of steps $k_0$ upper bounded by $n^2 + (n-1)m^2$ and then transmissions will be ceased.

Note that so far we considered the scenario where there is only one leading mass during every time step $k$ and it merges with only one nonzero mass variable every $n + m^2$ time steps. 
In other scenarios, we can consider multiple leading masses (i.e., the nonzero mass variables fulfill (\ref{great_z_prop1_det}) and (\ref{great_z_prop2_det}) for more than one node) which will speed up convergence since the follower masses will merge more frequently. \hspace*{\fill} $\square$
\end{pf}

The proof of Theorem~\ref{PROP1_max} analyzes the operation of Algorithm~\ref{algorithm_max} over a static and strongly connected digraph. 
It shows that the convergence time of the algorithm is equal to $n^2 + (n-1)m^2$ (where $n$ is the number of nodes and $m$ is the number of edges in the network). 
Note that the finite time \textit{deterministic} convergence of the algorithm is achievable due to the unique order $P_{lj}$ that each node $v_j$ assigns to its out-neighbors during the initialization steps. 
However, in most applications nowadays, such as control and coordination of autonomous vehicles or UAV swarms, we have that the structure of the network is time-varying. 
Analyzing the operation of Algorithm~\ref{algorithm_max} over time-varying digraphs is outside the scope of this paper and an extension of the unique order transmission strategy over time-varying networks is currently an open question. 
An extension of Algorithm~\ref{algorithm_max} over time-varying digraphs can be done through the analysis in \cite{2021:Rikos_Hadj_Splitting}. 
Specifically, \cite{2021:Rikos_Hadj_Splitting} presents an algorithm in which each node performs randomized transmissions towards its out-neighbors (i.e., it chooses an out-neighbor to perform a transmission according to a nonzero probability) and is shown to operate both in static and time-varying digraphs allowing convergence with high probability. 
Thus, an important future research direction is to extend the operation of Algorithm~\ref{algorithm_max} by applying a randomized transmission strategy. 
This extension will enhance the algorithm's convergence speed (see Fig.~$3$ and Fig.~$4$ in \cite{2021:Rikos_Hadj_Splitting}) and will allow convergence over time-varying digraphs; however, it will eliminate its \textit{deterministic} convergence characteristics, allowing instead finite time convergence with high probability.

\section{BOUNDING NUMBER OF TRANSMISSIONS AND COMPUTATIONS}\label{bound_trans_comp}

In this section we calculate an upper bound on the number of transmissions and the number of computations each node $v_j$ performs during Algorithm~\ref{algorithm_max}. 



\begin{theorem}\label{bound_trans}
During the operation of Algorithm~\ref{algorithm_max}, each node $v_j$ will perform at most $n + (n-1)m$ transmissions (where $n$ is the number of nodes and $m$ is the number of edges in the network) before quantized average consensus is reached and transmissions are ceased. 
\end{theorem}

\begin{pf}
During the operation of Algorithm~\ref{algorithm_max} every node $v_j$ performs (i) broadcast transmissions and (ii) directed transmissions. 
We provide an upper bound for each transmission type separately. 

\noindent
Broadcast Transmissions: 
It is easy to see that during Algorithm~\ref{algorithm_max} there are at most $n-1$ updates of the leading mass (see Theorem~\ref{PROP1_max}) which trigger broadcast transmissions. 
Considering also the broadcast transmission performed during the initialization procedure, we have that each node $v_j$ will perform at most $n$ broadcastings towards its out-neigbors during Algorithm~\ref{algorithm_max}. 

\noindent
Directed Transmissions (or Unicast Transmissions):
Our analysis is based on Proposition~$3$ in \cite{2014:RikosHadj} which provides a bound for the follower mass to travel via each edge in the digraph and thus necessarily also reach every other node. 
Specifically, considering a strongly connected digraph $\mathcal{G}_d = (\mathcal{V}, \mathcal{E})$ (with $n=|\mathcal{V}|$ nodes and $m=|\mathcal{E}|$ edges) we have that the number of iterations required for the follower mass to reach every node $v_i \in \mathcal{V}$ is bounded by $m^2$, where $m = | \mathcal{E} |$ is the number of edges. 
This result is derived from the fact that digraph $\mathcal{G}_d$ consists of $C_\beta$ cycles, where $C_\beta \in \{ 1, 2, ... m \}$ (see Proposition~$3$ in \cite{2014:RikosHadj}), and each cycle is traversed at most $m$ times by a follower mass until it reaches the node (say $v_i$) whose mass variables are the leading mass. 
Let us consider now that each node $v_j$ performs one directed transmission every time one follower mass traverses the cycle $C_{\beta_0}$ (to which node $v_j$ belongs to). 
This means that node $v_j$ will perform at most $m$ directed transmissions until this specific follower mass merges with the leading mass. 
Furthermore, we have that initially there are at most $n$ mass variables and during the operation of Algorithm~\ref{algorithm_max} there are at most $n-1$ mergings. 
As a result, $n-1$ follower masses will traverse the digraph in order to merge with the leading mass. 
This means that each node $v_j$ will perform at most $(n-1)m$ directed transmissions during Algorithm~\ref{algorithm_max}. 

Combining the results for (i) broadcast transmissions and (ii) directed transmissions, we have that during the operation of Algorithm~\ref{algorithm_max} each node $v_j$ will perform at most $n + (n-1)m$ transmissions before quantized average consensus is reached and transmissions are ceased. \hspace*{\fill} $\square$
\end{pf}




\begin{theorem}\label{bound_comp}
During Algorithm~\ref{algorithm_max}, each node $v_j$ will perform at most $1 + (n-1) (m + 1 + \mathcal{D}_{max}^-)$ computations (where $n$ is the number of nodes, $m$ is the number of edges and $\mathcal{D}_{max}^- = \max_{v_j \in \mathcal{V}} \mathcal{D}_j^-$) before quantized average consensus is reached and transmissions are ceased.
\end{theorem}

\begin{pf}
During the operation of Algorithm~\ref{algorithm_max}, we consider that a node $v_j$ performs computations if it executes (i) the Initialization steps or (ii) any of the Iteration Steps~$1$, $2$, $3$. 
The Initialization steps are executed only once. 
The Iteration steps are executed at time step $k$ only if a node $v_j$ receives a set of state variables $z^s_i[k]$, $y^s_i[k]$ during Iteration Step~$1$ or a set of mass variables $z_i[k]$, $y_i[k]$ during Iteration Step~$2$. 
Thus, the upper bound on computations is calculated according to the number of messages each node $v_j$ receives during Algorithm~\ref{algorithm_max}. 

Computations due to Received Mass Variables: 
We consider the cases of computations due to received mass variables that (i) fulfill Event-Triggered Conditions~$2$ of Algorithm~\ref{algorithm_max_1a} and (ii) fulfill Event-Triggered Conditions~$3$ of Algorithm~\ref{algorithm_max_1a}. 
\\ \noindent
For the first case we have that there are at most $n-1$ mergings of mass variables (see Theorem~\ref{PROP1_max}). 
This means that the amount of computations due to received mass variables that fulfill Event-Triggered Conditions~$2$ of Algorithm~\ref{algorithm_max_1a}, is upper bounded by $n-1$. 
For the second case we have that each node $v_j$ will perform a directed transmission for $(n-1)m$ times (see Theorem~\ref{bound_trans}). 
This directed transmission is the result of receiving a set of mass variables which fulfills Event-Triggered Conditions~$3$ of Algorithm~\ref{algorithm_max_1a}. 
This means that the amount of computations due to received mass variables that fulfill Event-Triggered Conditions~$3$ of Algorithm~\ref{algorithm_max_1a}, is upper bounded by $(n-1)m$. 
As a result, node $v_j$ will perform computations for received mass variables (by checking Event-Triggered Conditions~$2$ or Event-Triggered Conditions~$3$) for at most $(n-1)(m+1)$ times.

Computations due to Received State Variables: 
From Theorem~\ref{PROP1_max} we have that during Algorithm~\ref{algorithm_max} there are at most $n-1$ updates of the leading mass which trigger broadcast transmissions among nodes in the network. 
This means that node $v_j$ will receive a set of state variables $z^s_i[k]$, $y^s_i[k]$ from its in-neighbors for at most $(n-1)\mathcal{D}_{max}^-$ times, where $\mathcal{D}_{max}^- = \max_{v_j \in \mathcal{V}} \mathcal{D}_j^-$. 
As a result, node $v_j$ will perform computations for received state variables (by checking Event-Triggered Conditions~$1$) for at most $(n-1)\mathcal{D}_{max}^-$ times. 

We now consider the computations during the Initilization steps and we combine them with the results of computations for (i) received mass variables and (ii) received state variables. 
As a result, we have that during the operation of Algorithm~\ref{algorithm_max}, each node $v_j$ will perform at most $1 + (n-1) (m + 1 + \mathcal{D}_{max}^-)$ computations (where $\mathcal{D}_{max}^- = \max_{v_j \in \mathcal{V}} \mathcal{D}_j^-$) before quantized average consensus is reached, and computations along with transmissions are ceased. \hspace*{\fill} $\square$
\end{pf}

The result of Theorem~\ref{bound_comp} depends on the number of incoming messages since, from the operation of  Algorithm~\ref{algorithm_max}, each node performs a computation only after a transmission has been received. 
Furthermore, if no messages are received (i.e., no mass or state variables are received) then, during the operation of Algorithm~\ref{algorithm_max}, each node will not execute Iteration Steps~$1$, $2$ and $3$ and thus it will remain in hibernation mode (i.e., awaiting to receive signals without performing any computations). 
As a result, since the number of transmissions that each node performs during the operation of Algorithm~\ref{algorithm_max} is upper bounded (see Theorem~\ref{bound_trans}) then the number of computations is also upper bounded and the bound depends on the number of incoming messages.

\section{MEMORY AND ENERGY REQUIREMENTS FOR ACHIEVING QUANTIZED AVERAGE CONSENSUS}\label{energy_constr}

In this section we focus on the consumption of available resources from each node in the network. 
Specifically, we calculate an upper bound on the memory and energy each node $v_j$ requires during Algorithm~\ref{algorithm_max}.

\subsection{Required Memory}

We first characterize the memory requirements of each node $v_j$ during the operation of Algorithm~\ref{algorithm_max}.

\begin{prop}\label{Memory_prop}
During the operation of Algorithm~\ref{algorithm_max}, the memory requirement of each node $v_j$ is (i) $7 + 4 \mathcal{D}_{j}^-$ locations for integer values, and (ii) $ 2 + (3 + 2\mathcal{D}_{j}^-) \lceil \log_{2} n \rceil + (3 + 2\mathcal{D}_{j}^-) \lceil \log_{2} \sum_{j=1}^n | y_j[0] | \rceil $ bits for binary numbers. 
\end{prop}

\begin{pf}
During the operation of Algorithm~\ref{algorithm_max}, each node $v_j$ needs to store (i) $2$ transmission variables $S\_br_j$, $M\_tr_j$, (ii) $2$ mass variables $y_j[k]$, $z_j[k]$, (iii) $3$ state variables $y^s_j[k]$, $z^s_j[k]$, $q^s_j[k]$, and (iv) $4$ mass and state variables $y_i[k]$, $z_i[k]$, $y^s_i[k]$, $z^s_i[k]$ for each $v_i \in \mathcal{N}_j^-$ (i.e., $2$ mass variables and $2$ state variables that node $v_j$ may receive from each in-neighbor respectively). 
This means that the memory requirements of node $v_j$ is $7 + 4 \mathcal{D}_{j}^-$ locations for integer values (decimal numbers). 

In order to calculate the memory requirements for binary numbers we need to calculate the required bits for each one of the $7 + 4 \mathcal{D}_{j}^-$ integer values each node $v_j$ stores during the operation of the algorithm. Specifically, we have that node $v_j$ requires (i) $2$ bits for the binary transmission variables $S\_br_j$, $M\_tr_j$, (ii) $\lceil \log_{2} n \rceil$ and $\lceil \log_{2} \sum_{j=1}^n | y_j[0] | \rceil$ bits for the mass variables $z_j[k]$, $y_j[k]$ respectively, (iii) $\lceil \log_{2} n \rceil$, $\lceil \log_{2} \sum_{j=1}^n | y_j[0] | \rceil$ and $\lceil \log_{2} n \rceil + \lceil \log_{2} \sum_{j=1}^n | y_j[0] | \rceil$ bits for the state variables $z^s_j[k]$, $y^s_j[k]$, $q^s_j[k]$ respectively, and (iv) $2 ( \lceil \log_{2} n \rceil ) \mathcal{D}_{j}^-$ and $2 ( \lceil \log_{2} \sum_{j=1}^n | y_j[0] | \rceil ) \mathcal{D}_{j}^-$ bits for the mass and state variables $y_i[k]$, $z_i[k]$, $y^s_i[k]$, $z^s_i[k]$ for each $v_i \in \mathcal{N}_j^-$. 
Combining these $4$ cases, we have that the memory requirements of each node $v_j$ is $ 2 + (3 + 2\mathcal{D}_{j}^-) \lceil \log_{2} n \rceil + (3 + 2\mathcal{D}_{j}^-) \lceil \log_{2} \sum_{j=1}^n | y_j[0] | \rceil $ bits for binary numbers.  \hspace*{\fill} $\square$
\end{pf}

It is important to note here that during the operation of Algorithm~\ref{algorithm_max} the memory requirements of each node $v_j$ are greater than most algorithms in the current literature (e.g., \cite{2016:Chamie_Basar, 2011:Cai_Ishii} and references therein).
In most algorithms each node needs to store at most $2 + 2 \mathcal{D}_{j}^-$ integer values (i.e., at most $2$ integer values for the node's state and at most $2$ integer values for the state of each in-neighbor). 
However, in Algorithm~\ref{algorithm_max} each node $v_j$ needs to store $7 + 4 \mathcal{D}_{j}^-$ integer values in order to establish finite time convergence and transmission stopping. 
Specifically, each node $v_j$ needs to store (i) $2 + 2 \mathcal{D}_{j}^-$ integer values for mass variables, in order to establish finite time convergence, and (ii) $3 + 2 \mathcal{D}_{j}^-$ integer values for state variables along with $2$ integer values for transmission variables, in order to establish transmission stopping. 
This increase in the required memory of each node leads to preservation of the number of transmissions and the utilized bandwidth (since each node ceases transmissions once it achieves convergence to the quantized average). 
Considering that the energy cost for performing transmissions is much higher than the energy cost for performing computations \cite{2002:Chandrakasan}, this aspect of Algorithm~\ref{algorithm_max} is of particular importance since it leads to energy savings during the operation of each node. 
As a result, this characteristic makes Algorithm~\ref{algorithm_max} suitable for battery powered networks (as will be seen in the following section).

\subsection{Required Energy}

We now discuss the energy requirements of each node $v_j$ during the operation of Algorithm~\ref{algorithm_max}. 
Energy is consumed mainly during (i) communication, (ii) processing, and (iii) sensing. 
Therefore, before analyzing the operation of each node, we introduce the energy model from \cite{2002:Chandrakasan} which we will use to calculate the required energy per operation. 
\begin{list4}
\item[3. Sensing:] For each node $v_j$, the energy required to sense a bit is constant and equal to $\alpha_3$. The sensing power is 
\begin{equation}\label{eq_sense}
p_{\text{sense}} = \alpha_3 r ,
\end{equation}
for a sensing rate of $r$ bits/sec. A typical value of $\alpha_3$ is $50$ nJ/bit. 
\item [2. Processing:] For each node $v_j$, the energy required for aggregating $n_{\text{agg}}$ data streams into one stream is 
\begin{equation}\label{eq_process}
p_{\text{comp}} = \alpha_4 n_{\text{agg}} r ,
\end{equation}
where $r$ is the rate (bits/sec) and $\alpha_4$ is a constant (typically $5$ nJ/bit). 
\item[1. Communication:] For each node $v_j$, the energy required for transmitting to node $v_l$ is 
\begin{equation}\label{eq_trans}
p_{\text{trans}} = (\alpha_{11} + \alpha_{2} d(v_j, v_l)^n) r , 
\end{equation}
where $r$ is the rate (bits/sec), $d(v_j, v_l)$ is the distance between nodes $v_j$, $v_l$, $n$ is the path loss index, and $\alpha_{11}$,  $\alpha_{2}$ are constants (typically $45$ nJ/bit and $135$ nJ/bit, respectively). 
\end{list4}

For convenience we assume that during the operation of our algorithm, the above operations occur for $1$ second. 
We next analyze the required energy for each of the above operations separately. 
Then, the energy requirements of each node $v_j$ during the operation of Algorithm~\ref{algorithm_max} is the sum of these three results. 

\begin{lemma}\label{energy_receive}
During Algorithm~\ref{algorithm_max}, each node $v_j$ requires 
\begin{align}
& p^j_{\text{sense}} = & \nonumber \\ 
& \alpha_3 (m + 1 + \mathcal{D}_{max}^-) (n-1) (\lceil \log_{2} n \rceil + \lceil \log_{2} \sum_{j=1}^n | y_j[0] | \rceil) & \label{energy_receive_result}
\end{align}
nJ of energy for its sensing operation (i.e., for receiving values from its in-neighbors), where $n$ is the number of nodes, $m$ is the number of edges in the network, $\mathcal{D}_{max}^- = \max_{v_j \in \mathcal{V}} \mathcal{D}_j^-$, and $\alpha_3$ is decided by the specifications of the receiver node $v_j$ (typical value of $\alpha_3$ is $50$ nJ/bit). 
\end{lemma}

\begin{pf}
In order to calculate the required energy for each node $v_j$, we consider the analysis of Theorem~\ref{bound_comp} and Proposition~\ref{Memory_prop}. 
Specifically, we combine (i) the number of times each node $v_j$ received a set of mass and state variables (see Theorem~\ref{bound_comp}), (ii)  the number of bits the received sets of mass and state variables consist of (see Proposition~\ref{Memory_prop}), and (iii) the required energy for each node $v_j$ to sense a bit (shown in \eqref{eq_sense}). 

From Theorem~\ref{bound_comp}, we have that each node $v_j$ will receive a set of state variables $z^s_i[k]$, $y^s_i[k]$ from its in-neighbors for at most $(n-1)\mathcal{D}_{max}^-$ times. 
Furthermore, each node $v_j$ will receive a set of mass variables $z_i[k]$, $y_i[k]$ from its in-neighbors for at most $(n-1)(m+1)$ times. 
From Proposition~\ref{Memory_prop}, we have that each set of mass or state variables consists of at most $\lceil \log_{2} n \rceil + \lceil \log_{2} \sum_{j=1}^n | y_j[0] | \rceil$ bits. 
This means that for each node $v_j$, during the operation of Algorithm~\ref{algorithm_max}, the upper bound regarding the number of received bits is equal to $[(n-1)\mathcal{D}_{max}^- + (n-1)(m+1)] (\lceil \log_{2} n \rceil + \lceil \log_{2} \sum_{j=1}^n | y_j[0] | \rceil)$. 
Combining the upper bound regarding the number of received bits with the required energy to sense a bit in \eqref{eq_sense}, we have that each node $v_j$ requires $p^j_{\text{sense}}$ energy as in \eqref{energy_receive_result} for its sensing operation. \hspace*{\fill} $\square$
\end{pf}

\begin{lemma}\label{energy_process}
During Algorithm~\ref{algorithm_max}, each node $v_j$ requires 
\begin{align}
& p^j_{\text{comp}} = & \nonumber \\ 
& \alpha_4 [1 + 2(\mathcal{D}_{max}^-)^2] (n-1) (\lceil \log_{2} n \rceil + \lceil \log_{2} \sum_{j=1}^n | y_j[0] | \rceil) & \label{energy_process_result}
\end{align}
nJ of energy for its processing operation (i.e., for aggregating multiple streams into one stream), where where $n$ is the number of nodes, $m$ is the number of edges in the network, $\mathcal{D}_{max}^- = \max_{v_j \in \mathcal{V}} \mathcal{D}_j^-$, and $\alpha_4$ is decided by the specifications of the processing node $v_j$ (typical value of $\alpha_4$ is $5$ nJ/bit). 
\end{lemma}

\begin{pf}
In order to calculate the required energy for each node $v_j$, we consider the analysis of Theorem~\ref{bound_comp} and Proposition~\ref{Memory_prop} where we combine (i) the number of times each node $v_j$ received a set of mass and state variables, (ii) the number of bits the received sets of mass and state variables consist of, and (iii) the required energy for each node $v_j$ to aggregate multiple data streams into one stream (shown in \eqref{eq_process}). 

From Theorem~\ref{bound_comp}, we have that each node $v_j$ will receive a set of state variables $z^s_i[k]$, $y^s_i[k]$ from its in-neighbors for at most $(n-1)\mathcal{D}_{max}^-$ times. 
This means that node $v_j$ will have to aggregate $2 \mathcal{D}_{max}^-$ streams into two streams for at most $(n-1)\mathcal{D}_{max}^-$ times. 
Specifically, $v_j$ will have to aggregate $\mathcal{D}_{max}^-$ streams of $z^s_i[k]$, $v_i \in \mathcal{N}_j^-$, values into one stream $z^s_j[k+1]$ and $\mathcal{D}_{max}^-$ streams of $y^s_i[k]$, $v_i \in \mathcal{N}_j^-$, values into one stream $y^s_j[k+1]$. 
As a result, the required energy for node $v_j$ to process the received state variables is 
\begin{equation}\label{process_state}
p_1^j = \alpha_4 2(\mathcal{D}_{max}^-)^2 (n-1) (\lceil \log_{2} n \rceil + \lceil \log_{2} \sum_{j=1}^n | y_j[0] | \rceil) \ \text{nJ} .  
\end{equation}
Furthermore, each node $v_j$ will receive a set of mass variables $z_i[k]$, $y_i[k]$ from its in-neighbors for at most $(n-1)(m+1)$ times. 
However, the maximum number of aggregations of received mass variables $z_i[k]$, $y_i[k]$ is upper bounded by $n-1$ (i.e., there are at most $n-1$ aggregations of mass variables during the operation of Algorithm~\ref{algorithm_max}). 
This means that the required energy for node $v_j$ to process the received mass variables is 
\begin{equation}\label{process_mass}
p_2^j = \alpha_4 (n-1) (\lceil \log_{2} n \rceil + \lceil \log_{2} \sum_{j=1}^n | y_j[0] | \rceil) \ \text{nJ} .  
\end{equation}

As a result, combining \eqref{process_state} and \eqref{process_mass} we have that each node $v_j$ requires $p^j_{\text{comp}}$ energy as in \eqref{energy_process_result} for its processing operation. \hspace*{\fill} $\square$
\end{pf}

\begin{lemma}\label{energy_transmit}
During Algorithm~\ref{algorithm_max}, each node $v_j$ requires 
\begin{align}
& p^j_{\text{trans}} = & (n-1) (\alpha_{11} + \alpha_{2} d(v_j)^n) (m + 1) A & \label{energy_transmit_result}
\end{align}
nJ of energy for its transmission operation (i.e., for performing transmissions towards its out-neighbors), where 
$$
A = \lceil \log_{2} n \rceil + \lceil \log_{2} \sum_{j=1}^n | y_j[0] | \rceil , 
$$ 
and $n$ is the number of nodes, $m$ is the number of edges in the network, $d(v_j)$ is the distance between node $v_j$ and every $v_l \in \mathcal{N}_j^+$, $n$ is the path loss index, and $\alpha_{11}$, $\alpha_{2}$ are constants (typically $45$ nJ/bit and $135$ nJ/bit, respectively).  
[For notational simplicity, we assume that $d(v_j)^n = d(v_j, v_l)^n = d(v_j, v_{l'})^n$, for every $v_l, v_{l'} \in \mathcal{N}_j^+$.]
\end{lemma}

\begin{pf}
In order to calculate the required energy for each node $v_j$, we combine (i) the number of times each node $v_j$ performs a transmission of a set of mass and state variables (see Theorem~\ref{bound_trans}), (ii) the number of bits involved in the transmitted sets of mass and state variables (see Proposition~\ref{Memory_prop}), and (iii) the required energy for each node $v_j$ to perform a transmission towards its out-neighbors (shown in \eqref{eq_trans}). 

From Theorem~\ref{bound_trans}, we have that each node $v_j$ will transmit a set of state variables $z^s_j[k]$, $y^s_j[k]$ to its in-neighbors for at most $(n-1)$ times. 
Furthermore, each node $v_j$ will transmit a set of mass variables $z_j[k]$, $y_j[k]$ towards its in-neighbors for at most $(n-1) m$ times. 
From Proposition~\ref{Memory_prop}, we have that each set of mass or state variables consists of at most $\lceil \log_{2} n \rceil + \lceil \log_{2} \sum_{j=1}^n | y_j[0] | \rceil$ bits. 
This means that for each node $v_j$, during the operation of Algorithm~\ref{algorithm_max}, the upper bound regarding the number of transmitted bits is equal to $[(n-1)(m+1)] (\lceil \log_{2} n \rceil + \lceil \log_{2} \sum_{j=1}^n | y_j[0] | \rceil)$. 
Combining the upper bound regarding the number of transmitted bits with the required energy to transmit a bit in \eqref{eq_trans}, we have that each node $v_j$ requires $p^j_{\text{trans}}$ energy as in \eqref{energy_transmit_result} for its transmission operation. \hspace*{\fill} $\square$
\end{pf}

As a result, if we combine the above results, we obtain the total energy requirements of each node $v_j$ during the operation of Algorithm~\ref{algorithm_max}, which is equal to
\begin{align}
& p^j_{\text{total}} = & \nonumber \\ 
& (n-1) (\alpha_3 (m + 1 + \mathcal{D}_{max}^-) + \alpha_4 [1 + 2(\mathcal{D}_{max}^-)^2]) A \ + & \nonumber \\
& (n-1)(\alpha_{11} + \alpha_{2} d(v_j)^n) (m + 1) A \ \text{nJ}, & \label{energy_energy_result}
\end{align}
where  
$
A = \lceil \log_{2} n \rceil + \lceil \log_{2} \sum_{j=1}^n | y_j[0] | \rceil , 
$ 
$n$ is the number of nodes, $m$ is the number of edges, $\mathcal{D}_{max}^- = \max_{v_j \in \mathcal{V}} \mathcal{D}_j^-$, $\alpha_3$ is decided by the specifications of the receiver node $v_j$ (typical value of $\alpha_3$ is $50$ nJ/bit), $\alpha_4$ is decided by the specifications of the processing node $v_j$ (typical value of $\alpha_4$ is $5$ nJ/bit) and and $\alpha_{11}$, $\alpha_{2}$ are constants (typically $45$ nJ/bit and $135$ nJ/bit, respectively).

\begin{remark}
Note here that in \eqref{energy_energy_result}, we calculate the total required power of each node for the worst case scenario. 
The results in Section~\ref{bound_trans_comp} and Section~\ref{energy_constr} can also be extended for the cases where we want to (i) calculate the minimum required energy per node during the operation of Algorithm~\ref{algorithm_max}, and (ii) tune Algorithm~\ref{algorithm_max} to perform under an energy budget. 
In the first case, the analysis can consider the topology of the network, the distribution of the initial states of the nodes and the unique order each node assigns to its outgoing edges. 
In the second case, by utilizing the analysis of the first case, we can show how we can tune Algorithm~\ref{algorithm_max} to perform under an energy budget. 
Calculating the minimum required energy per node and adjusting the algorithm to perform under an energy budget are interesting directions for future research (but outside the scope of this paper).  $\hfill \blacksquare$
\end{remark}

\section{SIMULATION RESULTS}\label{results}

\begin{figure}[t]
\begin{center}
\includegraphics[width=1\columnwidth]{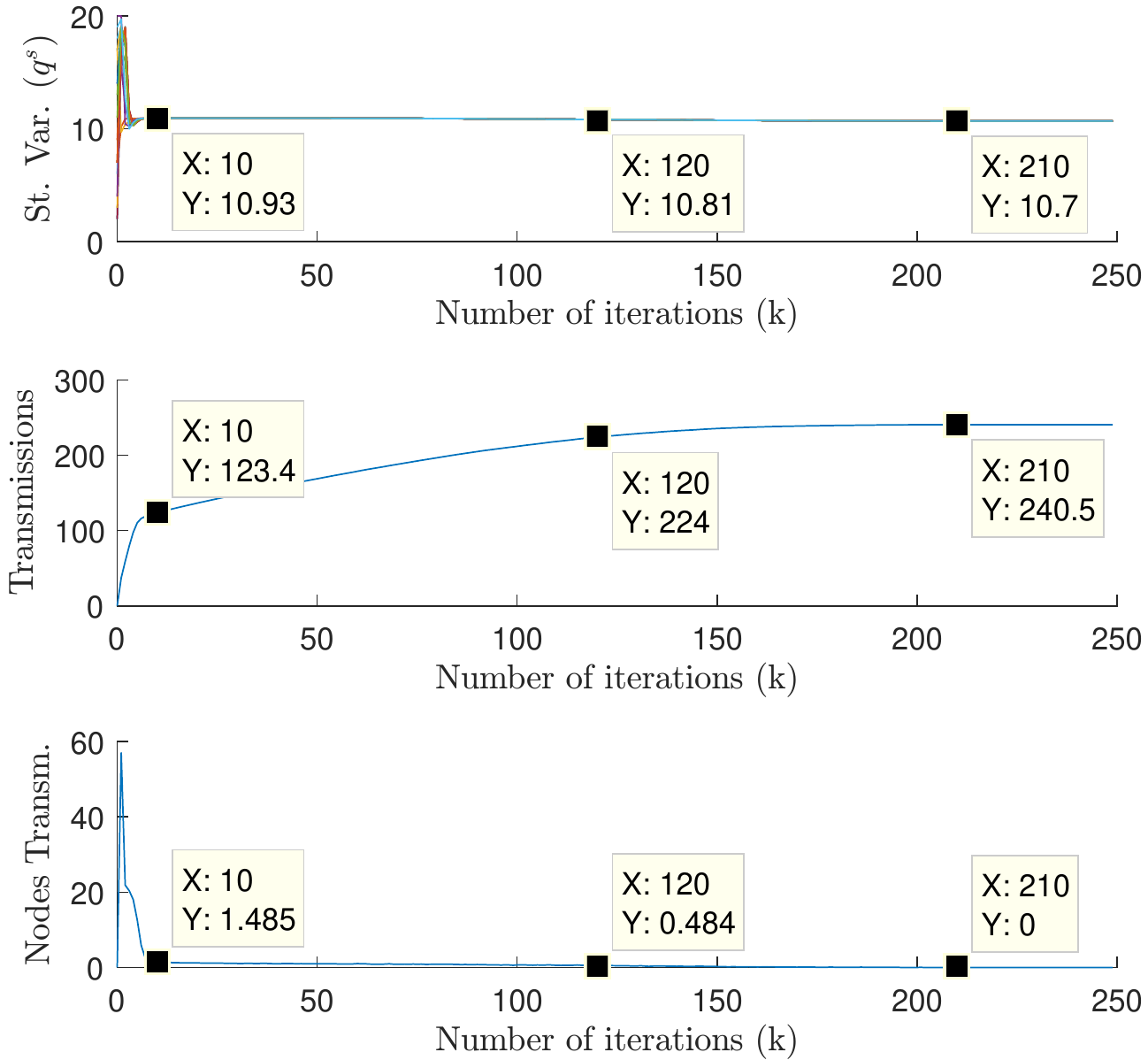}
\caption{Execution of Algorithm~\ref{algorithm_max} averaged over $1000$ random digraphs of $20$ nodes.  
\textit{Top figure:} Average values of node state variables plotted against the number of iterations (averaged over $1000$ random digraphs of $20$ nodes). \textit{Middle Figure:} Average accumulated number of transmissions plotted against the number of iterations (averaged over $1000$ random digraphs of $20$ nodes). \textit{Bottom Figure:} Average number of nodes performing transmissions plotted against the number of iterations (averaged over $1000$ random digraphs of $20$ nodes)\vspace{-0.2cm}.}
\label{fig_10_7}
\end{center}
\end{figure}

In this section, we illustrate the behavior of Algorithm~\ref{algorithm_max} and the advantages of its event triggered operation.
Specifically, for $1000$ randomly generated digraphs of $20$ nodes with identical (randomly chosen) integer initial values with average equal to $q = 214 / 20 = 10.7$, we show in Fig.~\ref{fig_10_7} (i) the average value of each node state variable at each time step, (ii) the average number of transmissions accumulated until each time step, and (iii) the average number of nodes performing transmissions at each time step. 
In Table~\ref{table_max_min_10_7}, we show the minimum, maximum and average values of the (i) total transmissions, and (ii) total required number of time steps for convergence, during the execution of Algorithm~\ref{algorithm_max} over these $1000$ randomly generated digraphs of $20$ nodes.

In Fig.~\ref{fig_10_7} it is interesting to notice the drop in the average number of nodes which perform transmissions at each time step (see bottom figure). 
Specifically, at time step $k=1$ we have that $57$ transmissions are performed because during initialization each node $v_j$ transmits its state variables and then, Event Trigger Conditions~$1$, $2$ and $3$ hold for some nodes in the digraph. 
However, the average number of transmissions at time step $k=10$ drops to $1.485$ and becomes almost equal to $1$ for time steps $k \geq 20$ which means that only one node performs transmissions after approximately $20$ time steps. 
Furthermore, we can see that for $k \geq 210$ the average number of transmissions becomes equal to zero (meaning that no node performs transmissions any more) since Event Trigger Conditions~$1$, $2$ and $3$ do not hold for any node. 
This means that every node has reached a common value, equal to $10.7$, which is equal to the average of the initial values (see top figure).
As a result, from Fig.~\ref{fig_10_7} we have that Algorithm~\ref{algorithm_max}, allows the nodes to reach quantized average consensus after an average number of $210$ time steps and $240.5$ transmissions.

From Table~\ref{table_max_min_10_7}, notice that the minimum number of transmissions is $103$ and the maximum number of transmissions is $368$ with the average being $240.547$. 
Furthermore, it is interesting to note that the minimum number of time steps for convergence is $5$, the maximum is $209$, with the average being $103.875$. 
Both results show that in practical scenarios (implemented over random directed graphs) the total number of transmissions and the total number of required time steps for convergence are much lower than the worst case upper bounds calculated in Section~\ref{bound_trans_comp} and Section~\ref{sec:Conv_analysis}, respectively.

\begin{table}[t]
\begin{center}
\captionof{table}{Minimum, Maximum, and Average Number of (i) Total Transmissions and (ii) Time Steps for Convergence, of Algorithm~\ref{algorithm_max} over $1000$ random digraphs of $20$ nodes.}
\label{table_max_min_10_7}
{\small 
\begin{tabular}{|c||c|c|c|}
\hline
 &\multicolumn{3}{c|}{$\#$ of Transmissions and Time Steps}\\
 &Min. & Max. & Average \\
\cline{1-4}
$\#$ of Transmissions & $103$ & $368$ & $240.547$\\
$\#$ of Time Steps & $5$ & $209$ & $103.875$\\
\hline
\end{tabular}
}
\end{center}
\end{table}
\vspace{0.2cm}


\section{CONCLUSIONS}\label{future}

In this work, we analyzed the quantized average consensus problem over wireless networks with nodes that are battery powered or utilize energy harvesting techniques. 
Quantized average consensus plays a key role in a number of applications, which aim at more efficient usage of network resources. 
We solved the quantized average consensus problem using a novel event-triggered distributed algorithm, which calculates the exact average (i.e., avoids the error introduced due to quantization) after a finite number of iterations, which we explicitly bounded. 
Furthermore, we showed that once the quantized average is calculated, transmissions are ceased from each node in the network. 
Then, we presented upper bounds on the number of transmissions and computations each node performs during the operation of the algorithm and used them to bound the memory and energy requirements of each node. 
Finally, we concluded with simulations which demonstrated the performance and the advantages of our algorithm. 
Note here that to the best of our knowledge, this is the first {\em deterministic} algorithm, which allows convergence to the exact quantized average of the initial values after a finite number of time steps without any specific requirements regarding the network that describes the underlying communication topology (see \cite{2016:Chamie_Basar}), while it achieves more efficient usage of available network resources due to its event-triggering operation and its transmission stopping capabilities.  

In the future, we plan to extend Algorithm~\ref{algorithm_max} to cases where (i) it performs under an energy budget, (ii) we have time-varying communication topologies, with bounded or unbounded transmission delays, and (iii) nodes aim to preserve the privacy of their initial states.

\vspace{-0.3cm}


\bibliographystyle{plain}        
\bibliography{bibliografia_consensus}           

\end{document}